\documentstyle[revtex]{aps}
\begin{document}

\begin{title} Viscosities of Quark-Gluon Plasmas \end{title}
\author{H. Heiselberg}
\begin{instit}
Nuclear Science Div., MS 70A-3307,
Lawrence Berkeley Laboratory, Berkeley, CA 94720, USA\\
\end{instit}
\date
\maketitle

\begin{abstract}
  The quark and gluon viscosities are calculated in quark-gluon plasmas to
  leading orders in the coupling constant by including screening. For weakly
  interaction QCD and QED plasmas dynamical screening of transverse
  interactions and Debye screening of longitudinal interactions controls the
  infrared divergences. For strongly interacting plasmas other screening
  mechanisms taken from lattice calculations are employed. By solving the
  Boltzmann equation for quarks and gluons including screening the viscosity is
  calculated to leading orders in the coupling constant. The leading
  logarithmic order is calculated exactly by a full variational treatment. The
  next to leading orders are found to be very important for sizable coupling
  constants as those relevant for the transport properties relevant for
  quark-gluon plasmas created in relativistic heavy ion collisions and the
  early universe.
\end{abstract}
\pacs{PACS numbers: 12.38.Bx, 97.60.Jd, 21.65.+f, 21.80.+a, 95.30.Cq}

\setlength{\oddsidemargin}{1.5cm}
\setlength{\evensidemargin}{\oddsidemargin}
\columnsep -.8cm
\twocolumn
\narrowtext

\section{INTRODUCTION}
\label{sec:intro}

Transport and relaxation properties of quark and gluon (QCD) plasmas are
important in a number of a different contexts.  They determine the time that it
takes a quark-gluon plasma formed in a heavy-ion collision to approach
equilibrium, and they are of interest in astrophysical situations such as the
early universe, and possibly neutron stars.

The basic difficulty in calculating transport properties of such plasmas, as
well as of relativistic electron-photon (QED) plasmas, is the singular nature
of the long-range interactions between constituents, which leads to divergences
in scattering cross sections similar to those for Rutherford scattering.  This
makes the problem of fundamental methodological interest, in addition to its
possible applications. The first approaches to describe the transport
properties of quark-gluon plasmas employed the relaxation time
approximation\cite{Ga,Ka,Da} for the collision term. This approximation
simplifies the collision integral enormously and transport coefficients are
related directly to the relaxation time. The latter is typically estimated from
a characteristic cross section times the density of scatterers.  In Refs.
\cite{Ka,Da} the divergent part of the total cross section at small momentum
transfers was assumed to be screened at momentum transfers less than the Debye
momentum. However, Debye screening influences only the longitudinal (electric)
part of the QED and QCD interactions, and the transverse (magnetic) part is
unscreened in the static limit.

Recently it has been shown that the physics responsible for cutting off
transverse interactions at small momenta is dynamical screening \cite{BP1}.
This effect is due to Landau damping of the exchanged gluons or photons.
Within perturbative QCD and QED rigorous analytical calculations of transport
coefficients to leading order have been made for temperatures high
\cite{BP1,BP3} as well as low \cite{deg} compared with the chemical potentials
of the constituents.

Transport processes depend on a characteristic relaxation time, $\tau_{tr}$,
of the particular transport process considered.
For example, in high temperature plasma the viscosities,
$\eta_i=w_i\tau_{\eta,i}/5$,
of particle type ${\em i}$ are proportional to the
characteristic times for viscous relaxation, $\tau_{\eta,i}\sim \tau_{tr}$,
which were first calculated in \cite{BP1} to leading order in the coupling
constant. More generally one finds that the typical transport relaxation
rates, that determines momentum stopping, thermal and viscous relaxation, is
in a weakly interacting QCD plasma
\begin{eqnarray}
   \frac{1}{\tau_{tr}} \propto \alpha_s^2\ln(1/\alpha_s) T
            \, + O(\alpha_s^2) \,  .  \label{tintro}
\end{eqnarray}
where the expansion is in terms of the fine structure constant
$\alpha_s=g^2/4\pi$.
The coefficients of proportionality to the leading order in $\alpha_s$
(in the following called the leading logarithmic order)
has been calculated analytically for
a number of transport processes in high temperature plasmas
\cite{BP1,BP3}. Likewise in a QED plasma
the typical transport relaxation rates for viscous processes, momentum
stopping, thermal and electrical conduction have the same dependence
as (\ref{tintro}) on the QED fine
structure constant $\alpha$ \cite{BP3}.

The dependence of the transport rates on the coupling constants is very
sensitive to the screening.  Besides the factor $\alpha_s^2$ from the matrix
element squared of the quark and gluon interactions, the very singular QCD
interactions for small momentum transfers lead to a logarithm,
$\ln(q_{max}/q_{min})$, of the maximum and minimum momentum transfers.  The
typical particle momenta limits the maximum momentum transfer, $q_{max}\sim T$,
and Debye and dynamical screening leads to effective screening for small
momentum transfers of order $q_{min}\sim q_D\sim gT$.  This gives the leading
logarithmic order in the coupling constant, $\ln(T/q_D)\sim\ln(1/\alpha_s)$, to
the transport rates (\ref{tintro}).

The calculations in \cite{BP1,BP3} were brief and dealt only with the leading
logarithmic order in the coupling constant with a given ansatz for the
distribution function. Here, more detailed calculations of the quark and gluon
viscosities in the high temperature quark-gluon plasmas are presented.  The
leading logarithmic order is calculated exactly by a variational method and the
next to leading order - the $\alpha_s^2$ term in (\ref{tintro}) - is calculated
as well.  Because $\alpha_s$ is not exponentially small, the next to leading
order is important in many realistic physical situations as relativistic heavy
ion collisions and the early universe.  Furthermore when the Debye screening
length is larger than the interparticle screening, which occur when
$\alpha_s\raisebox{-.5ex}{$\stackrel{>}{\sim}$} 0.1$ as we shall see below,
Debye and dynamical screening breaks down. Instead lattice gauge calculations
have found that quark-gluon plasmas seem to develop a constant screening mass,
$m_{pl}\simeq 1.1T$, for temperatures $T\raisebox{-.5ex}{$\stackrel{>}{\sim}$}
2-3 T_c$ and it is important to see what effects this alternative screening
mechanism has in strongly interacting plasmas.

We shall first describe in section II the transport theory we use, namely the
Boltzmann equation, and the screening of long range QCD and QED interactions.
In section III, we describe the process of shear flow and the variational
calculation necessary in order to find the viscosity.  In section IV we then
evaluate the collision term to leading logarithmic order with a simplifying
ansatz for the trial function and refer to Appendix A for a full and exact
variational calculation.  In section V we calculate the viscosity to
higher orders in the coupling constant and discuss strongly interacting
plasmas.  Finally, in section VI we give a summary and discuss generalizations
of the methods developed here to other transport coefficients.

\section{Transport Theory}

Transport processes are most easily described by the Boltzmann equation
\begin{eqnarray}
   (\frac{\partial}{\partial t} &+& {\bf v}_1\cdot\nabla_{\bf r}+
   {\bf F}\cdot\nabla_{{\bf p}_1} ) n_1
  \, = \,  2\pi\nu_2\sum_{234} |M|^2   \nonumber\\
 &\times& [n_1n_2(1\pm n_3)(1\pm n_4)-(1\pm n_1)(1\pm n_2)n_3n_4] \nonumber\\
 &\times&   \delta(\epsilon_1+\epsilon_2-\epsilon_3-\epsilon_4)
    \delta_{{\bf p}_1+{\bf p}_2;{\bf p}_3+{\bf p}_4}
   \, , \label{BE}
\end{eqnarray}
where $\epsilon_i$
is the energy and ${\bf p}_i$ the momentum of the quasiparticles,
${\bf F}$ some force acting on the quasi-particles, and the right hand side
of (\ref{BE}) is the collision term.
$n_i({\bf p}_i)$ are the Fermi and Bose quasi-particle distribution functions
for quarks and gluons and the signs $\pm$ include stimulated
emission and Pauli blocking.
The spin and color statistical factor $\nu_2$ is 16 for gluons and $12N_f$
for quarks and antiquarks with $N_f$ flavors.
$|M|^2$ is the squared matrix element for the
scattering process $12\to 34$, summed over final states and averaged over
initial states. It is related to the Lorentz-invariant matrix element
$|{\cal M}|^2$ by $|M|^2 =|{\cal M}|^2/
(16\epsilon_1\epsilon_2\epsilon_3\epsilon_4$).
For gluon-gluon scattering \cite{M} (see Fig. (\ref{bubble}))
\begin{eqnarray}
   |{\cal M}_{gg}|^2 &=& \frac{9}{4}g^4\left( 3-\frac{us}{t^2}-\frac{st}{u^2}
                      -\frac{ut}{s^2} \right)  \, ,\label{Mgg}
\end{eqnarray}
where $s$, $t$, and $u$ are the usual Mandelstam variables.
In Eq. (\ref{Mgg}) the  double counting of final states has been corrected
for by inserting a factor 1/2.
For quark-gluon scattering
\begin{eqnarray}
   |{\cal M}_{gq}|^2 = g^4(u^2+s^2)\left(\frac{1}{t^2}-\frac{4}{9us}\right),
   \label{Mgq}
\end{eqnarray}
and for scattering of two different quark flavors
\begin{eqnarray}
   |{\cal M}_{q_1q_2}|^2 = \frac{4}{9}g^4\frac{u^2+s^2}{t^2} \, .
   \label{Mqq}
\end{eqnarray}
The
matrix element for scattering of the same quark flavors or quark-antiquark
scattering is different at large momentum transfer but the same as (\ref{Mqq})
at small momentum transfers.

The $t^{-2}$ and $u^{-2}$ singularities in Eq. (\ref{Mgg}-\ref{Mqq})
lead to diverging transport cross sections and therefore
vanishing transport coefficients. Including screening, it was shown in
\cite{BP1,BP3,deg} that finite transport coefficients are obtained.
In fact, the leading contribution to transport coefficients comes from
these singularities. In the $t=\omega^2-q^2$ channel the singularity occurs
for small momentum ${\bf q}$ and energy $\omega$ transfers (see Fig.
(\ref{bubble})).

For small momentum transfer, $q\ll\epsilon_1,\epsilon_2\sim T$,
energy conservation implies that
$\omega=\epsilon_1-\epsilon_3\simeq {\bf v}_1\cdot {\bf q}=-{\bf v}_2\cdot
{\bf q}$
where ${\bf v}_i={\bf p}_i$. Therefore the
velocity projections transverse to ${\bf q}$ have lengths
$|{\bf v}_{1,T}|=|{\bf v}_{2,T}|=\sqrt{1-\mu^2}$, where $\mu=\omega/q$.
Consequently
${\bf v}_{1,T}\cdot{\bf v}_{2,T}=(1-\mu^2)\cos\phi$, where $\phi$ is the
angle between ${\bf v}_{1,T}$ and ${\bf v}_{2,T}$.
For $q\ll T$ we thus have
\begin{eqnarray}
   s \simeq -u &\simeq&  2p_1p_2(1-\cos\theta_{12}) \nonumber\\
    &\simeq& 2p_1p_2(1-\mu)(1-\cos\phi) \, ,
\end{eqnarray}
and the interactions splits into longitudinal and transverse ones, \cite{We}
\begin{eqnarray}
    \mid M_{gg}\mid^2
    = \frac{9}{8} g^4    \left|
   \frac{1}{q^2+\Pi_L}-\frac{(1-\mu^2)\cos\phi}{q^2-\omega^2+\Pi_T}\right|^2
   \, .   \label{Ms}
\end{eqnarray}
The interactions are modified by inclusion of the gluon, or photon,
self energies, $\Pi_L$ and $\Pi_T$ \cite{We} (see also Fig. (\ref{bubble}).
In the random-phase approximation the polarizations are
given in the long wavelength limit ($q\ll T$) by
\begin{eqnarray}
    \Pi_L(q,\omega) & = & q_{\rm D}^2
\left[1-\frac{\mu}{2}\ln\left(\frac{\mu+1}{\mu-1}\right)\right]  \, ,
\label{pl}\\
    \Pi_T(q,\omega)
     & = & q_{\rm D}^2 \left[\frac{\mu^2}{2}+\frac{\mu(1-\mu^2)}{4}
        \ln\left(\frac{\mu+1}{\mu-1}\right)\right] \, ,
\label{pt}
\end{eqnarray}
where $\mu=\omega/q$ and $q_{\rm D}=1/\lambda_{\rm D}$ is the Debye
wavenumber.
In a weakly-interacting high temperature  QCD plasma,\cite{We,KM}
\begin{eqnarray}
   q_D^2 = 4\pi(1+N_f/6)\alpha_s T^2 \, ,  \label{qD}
\end{eqnarray}
where $\alpha_s=g^2/4\pi$ is the fine structure
constant for strong interactions, the factor $(1+N_f/6)$ is the sum of
contributions from gluon screening, the ``1,'' and from light mass quarks, of
number of flavors, $N_f$.  In a high temperature
QED plasma, $q_{\rm D}^2=4\pi\alpha
T^2/3$, where $\alpha$ is the QED fine structure constant.

One should keep in mind that the self energies of (\ref{pl}) and (\ref{pt}) are
only valid in the long wavelength limit, i.e., for $q\ll T$. When $q\sim T$
other contributions of order $\alpha_s qT$ enter (see, e.g., \cite{We}) which
may be gauge dependent \cite{Ratata}. However, as long as $\alpha_s$ is small
all contributions from the self energies can be ignored in the gluon propagator
when $q\sim T$ because the matrix element squared already carry the order
$\alpha_s^2$.

In the above derivations we have consistently assumed that the screening was
provided in RPA by the gluon self energies which give the Debye and dynamical
screening of longitudinal and transverse interactions respectively.  Both
effects provide a natural effective cutoff of momentum transfer less than
$q_{min}\sim q_D$. These perturbative ideas must, however,
break down when the screening length becomes as short
as the interparticle spacing, i.e. $q_D\sim T$, or in terms of the
coupling constant
$\alpha_s\raisebox{-.5ex}{$\stackrel{<}{\sim}$} (4\pi(1+N_f/6))^{-1}\sim 0.1$
according to Eq. (\ref{qD}).
In lattice gauge calculations of quark-gluon plasma above twice
the temperature of the phase transition, $T_c\simeq 180$MeV, one finds strong
nonperturbative effects in the plasma so that the typical screening mass is
$m_{pl}\sim 1.1T$ \cite{Lattice}.  One may argue \cite{Shuryak} that
perturbation theory still applies for large momentum transfers so that the
matrix elements are given by the simple Feynman tree diagrams, but that
perturbation theory does not apply for small momentum transfers of order
$q\raisebox{-.5ex}{$\stackrel{<}{\sim}$} q_D$
and that one should rather insert the effective cutoff found by lattice
gauge calculations
\begin{eqnarray}
    \Pi_L \simeq \Pi_T \simeq m_{pl} \, ,\quad
   \alpha_s\raisebox{-.5ex}{$\stackrel{>}{\sim}$}
       0.1 \, . \label{Pim}
\end{eqnarray}
The phenomenological screening mass of (\ref{Pim}) provides
us with a method to extend our calculations of transport coefficients to
larger values for $\alpha_s$ and it can be combined to the
Debye and dynamical screening in weakly interacting quark-gluon
plasmas.

\section{The Viscosity}

With screening included in the interaction we can now proceed to
calculate transport properties as the viscosity.
In the presence of a small shear flow, ${\bf u}(y)$, in the $x$-direction
we obtain from the Boltzmann equation
\begin{eqnarray}
   & p_{1x}&v_{1y} \frac{\partial n_1}{\partial\epsilon_1}
                 \frac{\partial u_x}{\partial y}
    = 2\pi\nu_2\sum_{234} |M|^2  \nonumber\\
    &\times& \left[n_1n_2(1\pm n_3)(1\pm n_4)
   - (1\pm n_1)(1\pm n_2)n_3n_4 \right] \nonumber\\
   &\times&   \delta(\epsilon_1+\epsilon_2-\epsilon_3-\epsilon_4)
    \delta_{{\bf p}_1+{\bf p}_2;{\bf p}_3+{\bf p}_4}   \, . \label{BE2}
\end{eqnarray}
For small ${\bf u}$ we can furthermore
linearize the quasi-particle distribution function
\begin{eqnarray}
   n_i = n_i^{LE} +\frac{\partial n}{\partial\epsilon_p} \Phi_i
        \frac{\partial u_x}{\partial y}   \, , \label{n}
\end{eqnarray}
where the local equilibrium distribution function is
\begin{eqnarray}
   n_i^{LE} =(\exp[(\epsilon_i-{\bf u}\cdot{\bf p}_i)/T]\mp 1)^{-1},
     \label{LE}
\end{eqnarray}
and $\Phi_i$ is an unknown function that represents the deviations from
local equilibrium.
By symmetry $\Phi$ has to be on the form
\begin{eqnarray}
    \Phi = \hat{p}_x\hat{p}_y f(p/T) \, ,
\end{eqnarray}
where now the function $f$ must be determined from the Boltzmann equation.
Inserting (\ref{n} in the Boltzmann equation we find
\begin{eqnarray}
    p_{1x}v_{1y}\frac{\partial n_1}{\partial\epsilon_1}
    &=& 2\pi\nu_2\sum_{234} |M|^2
   [n_1n_2(1\pm n_3)(1\pm n_4) \nonumber\\
   && \hspace{15mm} -(1\pm n_1)(1\pm n_2)n_3n_4 ] \nonumber\\
   &\times&   \delta(\epsilon_1+\epsilon_2-\epsilon_3-\epsilon_4)
    \delta_{{\bf p}_1+{\bf p}_2;{\bf p}_3+{\bf p}_4}
   \nonumber\\ &\times&
   (\Phi_1+\Phi_2-\Phi_3-\Phi_4)
   \, . \label{BE3}
\end{eqnarray}
It is very convenient to define a scalar product of two real functions  by:
\begin{eqnarray}
   \langle\psi_1|\psi_2\rangle
   =-\nu_2\sum_{\bf p} \psi_1({\bf p})\psi_2({\bf p})
     \frac{\partial n}{\partial\epsilon_p} \,.
\end{eqnarray}
Thus Eq. (\ref{BE3} may be written on the form $|X\rangle=I|\Phi\rangle$
where $|X\rangle=p_xv_y$ and $I$ is the integral operator acting on
$\Phi$.
The viscosity is given in terms of $\Phi$ \cite{BPbook} and can now be
written as
\begin{eqnarray}
    \eta= -\nu_2\sum_{\bf p} p_xv_y \frac{\partial n}{\partial\epsilon_p}
          \Phi_{\bf p}
        = \langle X|\Phi\rangle \, . \label{eta}
\end{eqnarray}
Equivalently, the viscosity is given from (\ref{BE3} as
\begin{eqnarray}
    \eta = \frac{\langle X|\Phi\rangle^2}{\langle\Phi|I|\Phi\rangle} \, .
       \label{eta2}
\end{eqnarray}
Since $\langle .|. \rangle$ defines an inner product, the quantity
$\langle X|\Psi\rangle^2/\langle\Psi|I|\Psi\rangle$
is minimal for $\Psi=\Phi$ with the minimal value $\eta$. Equation (\ref{eta2}
is therefore convenient for variational treatment, which will be carried
out in Appendix A.

To find the viscosity we must solve the integral equation (\ref{BE3}
for which we have to evaluate
\begin{eqnarray}
    \langle\Phi|&I&|\Phi\rangle = 2\pi\nu_2\sum_{{\bf p}_1,{\bf p}_2,{\bf q}}
     |M|^2   n_1n_2(1\pm n_3) (1\pm n_4) \nonumber\\
     &\times& \frac{(\Phi_1+\Phi_2-\Phi_3-\Phi_4)^2}{4}
     \delta(\epsilon_1 +\epsilon_2 -\epsilon_3 -\epsilon_4 )  \, .  \label{PIP}
\end{eqnarray}
Momentum conservation requires that
${\bf p}_3={\bf p}_1+{\bf q}$ and ${\bf p}_4={\bf p}_2-{\bf q}$
where  ${\bf q}$ is the momentum transfer.
Introducing an auxilliary integral over energy transfers, $\omega$,
the delta-function in energy can be written
\begin{eqnarray}
     \delta(\epsilon_1 +\epsilon_2 -\epsilon_3 -\epsilon_4 ) &=& \int d\omega
   \frac{p_3}{p_1q}\delta(\cos\theta_1 -\mu-\frac{t}{2p_1q}) \nonumber\\
   &\times& \frac{p_4}{p_2q} \delta(\cos\theta_2 -\mu+\frac{t}{2p_2q})
    , \label{PIP2}
\end{eqnarray}
where $\theta_1$  is the polar angle between ${\bf q}$ and
${\bf p}_1$ and $\theta_2$ is the corresponding one between ${\bf q}$
and ${\bf p}_2$ (see Fig. 2).
Consequently, we find
\begin{eqnarray}
    \langle\Phi|I|\Phi\rangle &=& \frac{1}{4\pi^8T}
     \int_0^\infty dq\int_{-q}^q d\omega   \nonumber\\
   &\times&  \int_{(q-\omega)/2}^\infty dp_1\, p_1^2
              n_1(p_1)(1\pm n_1(p_1+\omega))
    \nonumber\\   &\times&
  \int_{(q+\omega)/2}^\infty dp_2\, p_2^2 n_2(p_2)(1\pm n_2(p_2-\omega))
   \nonumber\\   &\times&
  \int_0^{2\pi}\frac{d\phi}{2\pi}
      |M|^2  (\Phi_1+\Phi_2-\Phi_3-\Phi_4)^2
    \, . \label{PIP3}
\end{eqnarray}

This integral equation for $\Phi$ has been solved in a few cases under
simplifying circumstances. F.ex., in Fermi liquids the sharp Fermi surface
restricts all particle momenta near the Fermi surface and with a simplified
form for the scattering matrix element techniques have been developed to
calculate a number of transport coefficients exactly \cite{BPbook}. For the QCD
and QED plasmas the very singular interaction can, once screened, be exploited
since it allows an expansion at small momentum transfers. Thus an analytical
calculation of the transport coefficients can be carried out at least to
leading logarithmic order in the coupling constant \cite{BP1,BP3,deg}.

\section{Viscosity to Leading Logarithmic Order}

In Ref. \cite{BP1}, through solution of the Boltzmann kinetic
equation, the first viscosity of a quark-gluon plasma was derived
to leading logarithmic order in the QCD coupling
strength.  We will in the following give a more thorough and exact
derivation of the quark and gluon viscosity. The total viscosity,
to leading order, is an additive sum of the gluon and quark
viscosities, $\eta=\eta_g+\eta_q$.

The leading logarithmic order comes from small momentum transfers because of
the very singular matrix element (\ref{Ms}) dominates. For small $q$ the
kinematics simplify enormously and, as we will now show, the integrals
separate allowing almost analytical calculations.
First, we can set the lower limits on the $p_1$ and $p_2$ integrals to zero,
however, then replacing the upper limit on $q$ by the natural cutoff from
the distribution functions which is $q_{max}\sim T$.
Thus we find from (\ref{PIP3})
\begin{eqnarray}
    \langle\Phi|I|\Phi\rangle &=& \frac{1}{2\pi^8T}
   \int_0^\infty dp_1\, p_1^2n_1(1\pm n_1)   \nonumber\\
   &\times&   \int_0^\infty dp_2\, p_2^2n_2(1\pm n_2) \nonumber\\
   &\times&   \int_0^{q_{max}} qdq \int_{-1}^1 \frac{d\mu}{2}
   \int_0^{2\pi}\frac{d\phi}{2\pi}  \nonumber\\
    &\times&   |M|^2
   (\Phi_1+\Phi_2-\Phi_3-\Phi_4)^2    \, . \label{I3}
\end{eqnarray}
to leading logarithmic order

The solution to the integral equation or equivalently the variational
calculation of (\ref{eta2}) is quite technical and is for that reason given
in Appendix A. A much simpler calculation is to make the
standard assumption in viscous processes, i.e., to take the
trial function as
\begin{eqnarray}
    f(p/T) = (p/T)^2 \, . \label{fa}
\end{eqnarray}
As will be shown in Appendix A this turns out to be a very good
approximation. It is accurate to more than 99\% for reasons also explained
in the appendix. $f$ can be defined up to any constant which cancels
in (\ref{eta2}) and therefore never enters in the viscosity.

The quantity $(\Phi_1+\Phi_2-\Phi_3-\Phi_4)^2$ can be averaged over $x$- and
$y$-directions while keeping $\mu$ and $\phi$ fixed. This corresponds to
keeping the relative positions of the three vectors ${\bf q}$, ${\bf p}$,
and ${\bf p}'$ fixed relative to each other
and rotating this system over the three Euler
angles (see also Appendix A). Consequently, we obtain
\begin{eqnarray}
  \langle(\Phi_1+\Phi_2&-&\Phi_3-\Phi_4)^2\rangle = \frac{q^2}{15T^4}
      [3({\bf p}_2-{\bf p}_1)^2 \nonumber\\
  && \hspace{2cm}  +({\bf {\hat q}}\cdot({\bf p}_2-{\bf p}_1))^2  ]
      \nonumber\\
  &=& \frac{q^2}{15T^4} [(3+\mu^2)(p_1^2+p_2^2) \nonumber\\
  && \hspace{5mm} -2p_1p_2(4\mu^2+3(1-\mu^2)\cos\phi) ]   \, . \label{p2}
\end{eqnarray}
The integrals over $p_1$ and $p_2$ in (\ref{I3}) are elementary.
Next we perform the integrations or averages over $\mu$ and $\phi$
required in (\ref{I3}).
We note in passing that the term in (\ref{p2}) proportional to $p_1p_2$
vanishes and that $\mu^2$ effectively can be replaced by 1/3
(see Appendix A).
Let us first consider the case of gluon-gluon
scattering inserting $|M_{gg}|^2$ from (\ref{Mgg}). We thus find
\begin{eqnarray}
   \langle\Phi|I|\Phi\rangle &=& \frac{2^8\pi^3}{45} \alpha_s^2 T^3
     \int_0^{q_{max}} q^3 dq \int_0^1 d\mu  \nonumber\\
   && [ \frac{1}{|q^2+\Pi_L(\mu)|^2}
    +\frac{1/2}{|q^2+\Pi_T(\mu)/(1-\mu^2)|^2} ]  .\nonumber\\
   &&  \label{I4}
\end{eqnarray}
This integral is discussed in detail in appendix B. For the longitudinal
interactions $\Pi_L\simeq q_D^2$ due to Debye screening and the leading term
is a logarithm of the ratio of maximum to minimum momentum transfer,
$\ln(q_{max}/q_{min})\sim\ln(T/q_D)$. Likewise for the transverse interactions
$\Pi_T\simeq i(\pi/4)\mu q_D^2$ due to Landau damping and the dependence on
$\mu=\omega/q$ provides sufficient screening to render the integral finite
and the leading term is the same logarithm as for the longitudinal
interactions.
Whereas the  details of the screening are unimportant for the leading
logarithmic order, they are important for the higher orders and they are
calculated in detail in appendix B.
The final result is thus to leading logarithmic order
\begin{eqnarray}
    \langle\Phi|I|\Phi\rangle = \frac{2^7\pi^3}{15}
     \alpha_s^2\ln(T/q_D) T^3  \, ,     \label{I5}
\end{eqnarray}
Since $\langle\Phi|X\rangle= (64\xi(5)/\pi^2)T^3$, we find from
(\ref{eta2}) and (\ref{I5})
\begin{eqnarray}
    \eta_{gg} &=& \frac{2^5 15\xi(5)^2}{\pi^7} \frac{T^3}{\alpha^2_s\ln(T/q_D)}
   \nonumber\\
   &\simeq& 0.342 \frac{T^3}{\alpha_s^2\ln(1/\alpha_s)} \, , \label{etagg}
\end{eqnarray}
to leading logarithmic order in $\alpha_s=g^2/4\pi$.

To obtain the full gluon viscosity we must add
scattering on quarks and antiquarks which is calculated analogously
and only has a few factors different. Firstly, from (\ref{Mgg})
and (\ref{Mgq}) we see that
the matrix element squared is a factor 4/9 smaller. Secondly,
the statistical factor is $\nu_2=12N_f$ instead of 16. Thirdly, in
integrating over the factor $(p_1^2+p_2^2)$ in Eq. (\ref{p2}) we note that
the distribution function, $n_2$, in Eq. (\ref{I3}) is now a fermion one.
Consequently, the
$p_1$ and $p_2$ integrations give a factor $(1/2+7/8)/2$ less for
gluon-quark collisions as compared to gluon-gluon collisions and we find
\begin{eqnarray}
   \eta_g = (\eta_{gg}^{-1}+\eta_{gq}^{-1})^{-1}
          &=& \frac{\eta_{gg}}{1+11N_f/48} .
\end{eqnarray}
In \cite{BP1} the slightly different result
$\eta_g = \eta_{gg}/(1+N_f/6)$ was obtained.

The quark viscosity can be obtained analogously to the gluon one.
The quark viscosity due to collisions on quarks only, $\eta_{qq}$,
deviates from $\eta_{gg}$ by a factor $(4/9)^2$ in the matrix elements
and differences in having Fermi and Bose integrals. By comparing to
(\ref{eta2},\ref{eta},\ref{PIP}) we find
\begin{eqnarray}
   \eta_{qq} &=& \eta_{gg} \frac{(15/16)^2}{(4/9)^2(7/8)(1/2)}
            \, = \eta_{gg} \frac{5^23^6}{2^8 7} \, .
\end{eqnarray}
Note that the statistical factors $\nu$
cancel in $\eta_{gg}$ and $\eta_{qq}$.
Including quark scatterings on gluons lead to similar factors in
in $\langle \Phi|I|\Phi\rangle$, namely a factor (9/4) from the matrix element,
a factor $16/12N_f$ from statistics, and a factor $(8/7+2)/2$ from Bose
instead of Fermi integrals. Thus
\begin{eqnarray}
   \eta_q = \frac{\eta_{qq}}{1+33/7N_f}
   \simeq 2.2 \frac{1+11N_f/48}{1+7N_f/33} N_f \eta_g \, ,
\end{eqnarray}
which for $N_q=2$ results in $\eta_q=4.4\eta_g$, a quark viscosity that
is larger than the gluon one partly because the gluons generally
interact stronger than the quarks and partly because of differences between
Bose and Fermi distribution functions.

\section{Viscosity to higher orders in $\alpha_s$}

The leading logarithmic order dominates
at extremely high temperatures, where the running coupling constant is small,
but it is insufficient at lower temperatures.
The next to leading order correction to the viscous rate
in the coupling constant is of order $\alpha_s^2$. It may be
significant because the leading logarithm is a slowly increasing function.
In the derivation of the leading logarithmic order, Eq. (\ref{I5}),
we have been very cavalier with any factors entering in the logarithm,
which are of order $\alpha_s^2$.
It was only argued that the leading logarithmic order
$\ln(q_{max}/q_{min})\sim\ln(T/q_D)$  because $q_{max}$ and $q_{min}$
were of order $\sim T$ and $\sim q_D$ respectively.
Finally, if thermal quark-gluon plasmas
are created in relativistic heavy ion
collisions at CERN and RHIC energies, the temperatures achieved will
probably be below a GeV. We can thus estimate the interaction strength
from the running coupling constant
$\alpha_s\simeq 6\pi/(33-2N_f)\ln(T/\Lambda)$ which, with
$\Lambda\simeq 150$MeV and $T\raisebox{-.5ex}{$\stackrel{<}{\sim}$} 1$GeV,
gives $\alpha_s\raisebox{-.5ex}{$\stackrel{>}{\sim}$} 0.4$.
For such large coupling constants Debye and dynamical screening is replaced
by an effective screening mass, $m_{pl}$, as discussed above which will
affect the viscosity considerably.

To calculate the viscosity to order $\alpha_s^2$ exactly, the 5-dimensional
integral of (\ref{PIP3}) must be evaluated numerically and at the same time a
variational calculation of $\Phi$ must be performed.  This is a very difficult
task and we shall instead use the information obtained in the previous section,
that the trial function $f\propto p^2$ is expected to be an extremely good
approximation.  With that ansatz for the trial function, it is then straight
forward to calculate the integral of (\ref{PIP2}) numerically and find the
viscosity to order $\alpha_s^2$ for the given screening mechanism.
The 5-dimensional numerical evaluation of the
collision integral of (\ref{PIP2}) is a complicated function of the coupling
constant. It is convenient to write it in terms of the function Q
\begin{eqnarray}
    \langle\Phi|I|\Phi\rangle_{gg} &=&
    \frac{2^7\pi^3}{15} \alpha_s^2 Q\left(\frac{q_{max}^{(gg)}}{q_{min}}\right)
    T^3 \, , \label{IQ7}
\end{eqnarray}
where the index {\em gg} refers to gluon-gluon scattering but the analogous
definitions applies to gluon-quark and quark-quark scattering.
The function $Q$ and the effective maximum and minimum momentum
transfer, $q_{max}$ and $q_{min}$, are given in Appendix B.
In weakly interacting plasmas, where the screening is provided by Debye
and dynamical screening, the function Q is basically just a
logarithm of the ratio of the maximum and minimum momentum transfer, i.e.,
\begin{eqnarray}
   Q\left(\frac{q_{max}}{q_{min}}\right)
   = \ln\left(\frac{q_{max}}{q_{min}}\right) \, , \quad
\alpha_s\raisebox{-.5ex}{$\stackrel{<}{\sim}$} 0.1
\end{eqnarray}
By numerical integration we find that the distribution functions leads to an
effective cutoff of $q_{max}^{(gg)}\sim 3T$.  This is because the distribution
functions are weighted with several powers of particle momenta and thus
contribute the most for $p\simeq 3T$. The effective cutoff is slightly
larger for quark-gluon and quark-quark scattering because the Fermi
distribution functions emphasize large momenta than the Bose ones.
Debye and dynamical screening leads to
$q_{min}\simeq 1.26q_D$ as described in Eq. (\ref{qmin}) and so from
(\ref{QDD})
\begin{eqnarray}
   Q\left(\frac{q_{max}^{(gg)}}{q_{min}}\right) &=&
    \ln\left(\frac{0.44}{\alpha_s(1+N_f/6)}\right)
    \, , \quad \alpha_s\raisebox{-.5ex}{$\stackrel{<}{\sim}$} 0.1 \, .
\label{I7}
\end{eqnarray}
The numerical factor inside the logarithm, which gives the order
$\alpha_s^2$, is discussed in more detail in Appendix B.

In the other limit, $q_D\raisebox{-.5ex}{$\stackrel{>}{\sim}$} T$ or
equivalently $\alpha_s\raisebox{-.5ex}{$\stackrel{>}{\sim}$} 0.1$,
perturbative ideas breaks down and we assume
an effective screening mass taken from lattice calculations,
$q_{min}\simeq 1.1T$, as described by Eq. (\ref{Pim}).
Thus we find (see (\ref{Qmpl}))
\begin{eqnarray}
   Q(q_{max}^{(gg)}/q_{min})= 0.626 \, , \quad
\alpha_s\raisebox{-.5ex}{$\stackrel{>}{\sim}$} 0.1 \, ,
\end{eqnarray}
and similarly for quark-gluon and gluon-gluon scattering
$Q(q_{max}^{(gq)}/q_{min})= 0.819$ and
$Q(q_{max}^{(qq)}/q_{min})= 1.024$ respectively.

Adding gluon-gluon and gluon-quark scatterings we obtain the gluon viscosities
\begin{eqnarray}
    \eta_g &=& \frac{2^515\xi(5)^2}{\pi^7} \, \frac{T^3}{\alpha_s^2}
       \nonumber\\ && [ Q\left(\frac{q^{(gg)}_{max}}{q_D}\right)
     +  \frac{11N_f}{48} Q\left(\frac{q^{(gq)}_{max}}{q_D}\right)
        ]^{-1}                    \, ,    \label{eg}
\end{eqnarray}
which extends Eq. (\ref{etag}) to higher orders.
In weakly interacting plasmas (\ref{eg}) reduces to
\begin{eqnarray}
     \eta_g  &\simeq& 0.342\, \frac{T^3}{\alpha_s^2}
          [ \ln\left(\frac{0.44}{\alpha_s(1+N_f/6)}\right)    \nonumber\\
    &&\hspace{.1cm} +
    \frac{11N_f}{48}\ln\left(\frac{0.72}{\alpha_s(1+N_f/6)}\right) ]^{-1}
           \, , \alpha_s\raisebox{-.5ex}{$\stackrel{<}{\sim}$} 0.1 \label{etag}
\end{eqnarray}
to leading orders in $\alpha_s$.
In strongly interacting plasmas we obtain by inserting (\ref{Qmpl}) in
(\ref{IQ7})
\begin{eqnarray}
     \eta_g  &\simeq& 0.55\, \frac{T^3}{\alpha_s^2}
          \left[ 1 + 1.31 \frac{11N_f}{48} \right]^{-1}
           \, , \, \alpha_s\raisebox{-.5ex}{$\stackrel{>}{\sim}$} 0.1
\label{etag1}
\end{eqnarray}

In Fig. (\ref{figetag})
we show the gluon viscosity with the various assumptions
for screening. With dash-dotted curve the result of Eq. (\ref{etag1})
assuming a constant screening mass, $m_{pl}=1.1T$, is shown. With dashed
curve the numerical result assuming Debye and dynamical screening of Eqs.
(\ref{pl}) and (\ref{pt}) is shown.
For $\alpha_s\raisebox{-.5ex}{$\stackrel{<}{\sim}$} 0.05$ it is
given by Eq. (\ref{etag}) to a good approximation whereas for
$\alpha_s\raisebox{-.5ex}{$\stackrel{>}{\sim}$} 0.05$  the result of
Eq. (\ref{qin}) is better.
The final viscosity shown by full curve is obtained by
combining the two limits, i.e., applying Debye and dynamical screening
in weakly interacting plasmas when $q_D\raisebox{-.5ex}{$\stackrel{<}{\sim}$}
T$ or equivalently
$\alpha_s\raisebox{-.5ex}{$\stackrel{<}{\sim}$} 0.1$ but an effective
screening mass $m_{pl}=1.1T$
as given by Eq. (\ref{Pim}) when
$\alpha_s\raisebox{-.5ex}{$\stackrel{>}{\sim}$} 0.1$. This corresponds to
chosing the smallest value of the viscosities as seen in Fig. (\ref{figetag}),
i.e., the two limits of Eqs. (\ref{etag}) and (\ref{etag1}).

Similarly, adding quark-quark and quark-gluon scatterings we find the quark
viscosity
\begin{eqnarray}
  \eta_q  &=&  \frac{5^33^6\xi(5)^2}{2^311\pi^7} N_f \frac{T^3}{\alpha_s^2}
     \nonumber\\  &&    [ Q\left(\frac{q^{(gq)}_{max}}{q_D}\right)
      +   \frac{7N_f}{33} Q\left(\frac{q^{(qq)}_{max}}{q_D}\right)
          ]^{-1}     \, ,\label{eq}
\end{eqnarray}
which in weakly interacting plasmas gives
\begin{eqnarray}
   \eta_q  &\simeq&  0.752 N_f \frac{T^3}{\alpha_s^2}
        \, [\ln\left(\frac{0.72}{\alpha_s(1+N_f/6)}\right) \nonumber\\
   &&\hspace{5mm}
   +\frac{7N_f}{33}\ln\left(\frac{1.15}{\alpha_s(1+N_f/6)}\right)]^{-1}
           \, , \alpha_s\raisebox{-.5ex}{$\stackrel{<}{\sim}$} 0.1 ,
\label{etaq}
\end{eqnarray}
and in the strongly interacting plasmas
\begin{eqnarray}
   \eta_q  &\simeq&  0.92 N_f \frac{T^3}{\alpha_s^2}
        \left[ 1 + 1.25\frac{7N_f}{33} \right]^{-1}
           \, , \alpha_s\raisebox{-.5ex}{$\stackrel{>}{\sim}$} 0.1 \, .
\label{etaq1}
\end{eqnarray}

The quark viscosity increases with the number of quark flavors, $N_f$, whereas
the gluon viscosity decreases as can be seen in Fig.  (\ref{figeta2}) and
(\ref{figeta3}), where the viscosities are shown for two and three flavors
respectively.  The total viscosity of a quark-gluon plasma,
$\eta=\eta_g+\eta_q$, is dominated by the quark viscosity.

{}From the definition of the viscosity in terms of the collision integral
(\ref{eta}) and (\ref{PIP}), which only contains positive quantities, it
follows trivially that the viscosity is positive as is a physical necessity.
The resulting viscosities of Eqs. (\ref{eg}) and (\ref{eq}) are positive
quantities whereas the $\alpha_s\raisebox{-.5ex}{$\stackrel{<}{\sim}$} 0.1$
expansions of Eqs. (\ref{etag}) and
(\ref{etaq}) are not when extended to the region
$\alpha_s\raisebox{-.5ex}{$\stackrel{>}{\sim}$} 0.5$. This
explains the results found in \cite{Thoma}, where it was claimed that estimates
of the next to leading order $\alpha_s^2$ could lead to a negative viscosity.

Contributions from vertex corrections should also be considered. In fact for
the calculation of the quasiparticle damping rate, $\gamma_p$, Braaten and
Pisarski \cite{Pisarski} found that vertex corrections contributed to leading
order $\gamma^{(g)}_{p=0}\simeq 6.6\alpha_s$ for zero gluon momenta, $p$.
Vertex corrections do also contribute to order $\alpha_s$ for large
quasiparticle momenta, $p\gg gT$, but they can here be ignored since the
leading order is $\gamma^{(g)}_p=3\alpha_s\ln(1/\alpha_s)$ as explained in
\cite{HP}. For the viscosity, however, vertex corrections can be ignored since
the extra vertices adds a factor $\alpha_s^2$. Even though integration over
soft momenta may cancel a factor $\alpha_s$ the result is still of higher order
in the coupling constant.

Writing each of the viscosities $\eta_i$ $(i=q,g)$ in terms of the viscous
relaxation time, $\tau_{\eta i}$, as
\begin{eqnarray}
    \eta_i = w_i\tau_{\eta i}/5,   \label{etai}
\end{eqnarray}
where $w_g=(32\pi^2/45)T^4$ and $w_q=(N_f7\pi^2/15)T^4$
are the gluon and quark enthalpies
respectively, we obtain the viscous relaxation rate for gluons
\begin{eqnarray}
    \frac{1}{\tau_{\eta,g}} &=& \frac{\pi^9}{3^35^3\xi(5)^2} \,
   \frac{T}{\alpha_s^2} \nonumber\\ &\times&
    \left[Q\left(\frac{q_{max}^{(gg)}}{q_{min}}\right)
    +{11{N_f}\over 48}Q\left(\frac{q_{max}^{(gg)}}{q_{min}}\right)\right]
    ,\quad \alpha_s^2  T  , \label{tg}
\end{eqnarray}
and quarks and antiquarks
\begin{eqnarray}
     \frac{1}{\tau_{\eta,q}} &=& \frac{11\pi^92^37}{3^75^5\xi(5)^2}\,
   \frac{T}{\alpha_s^2} \nonumber\\ &\times&
    \left[Q\left(\frac{q_{max}^{(gq)}}{q_{min}}\right)
    +\frac{7N_f}{33}Q\left(\frac{q_{max}^{(qq)}}{q_{min}}\right) \right]
    ,\quad   \alpha_s^2 T \, .  \label{tq}
\end{eqnarray}
The viscous relaxation times, $\tau_{\eta,g}$, $\tau_{\eta,q}$ and
$\tau_{\eta}=1/(\tau^{-1}_{\eta,g}+\tau^{-1}_{\eta,q})$
are thus very similar to
the corresponding viscosities when divided by a factor of $T^4$. The curves on
Figs.  (\ref{figeta2}) and (\ref{figeta3}) therefore applies to the viscous
relaxation times (times temperature) as well when divided a factor of $\sim
1.4$ and $\sim 0.92N_f$ for gluons and quarks respectively according to Eq.
(\ref{etai}).

In weakly interacting plasma the viscous rates can be approximated by
\begin{eqnarray}
    \frac{1}{\tau_{\eta,g}}  &\simeq& 4.11\,
 \alpha_s^2
          [ \ln\left(\frac{0.44}{\alpha_s(1+N_f/6)}\right)    \nonumber\\
    &&\hspace{.1cm} +
    \frac{11N_f}{48}\ln\left(\frac{0.72}{\alpha_s(1+N_f/6)}\right) ]
          , \quad  \alpha_s\raisebox{-.5ex}{$\stackrel{<}{\sim}$} 0.1
\end{eqnarray}
and
\begin{eqnarray}
     \frac{1}{\tau_{\eta,q}} &\simeq& 1.27\,  \alpha_s^2
        \, [\ln\left(\frac{0.72}{\alpha_s(1+N_f/6)}\right) \nonumber\\
   &&\hspace{5mm}
   +\frac{7N_f}{33}\ln\left(\frac{1.15}{\alpha_s(1+N_f/6)}\right)]
          , \, \alpha_s\raisebox{-.5ex}{$\stackrel{<}{\sim}$} 0.1  \, ,
\end{eqnarray}
to leading orders in $\alpha_s$.

\section{Summary}

By solving the Boltzmann equation for quarks and gluons the viscosities in
quark-gluon plasmas were calculated to leading orders in the coupling constant.
Inclusion of dynamical screening of transverse interactions, which controls the
infrared divergences in QED and QCD, is essential for obtaining finite
transport coefficients in the weakly interacting plasmas.  The solution of the
transport process was extended to strongly interacting plasmas by assuming an
effective screening mass of order $m_{pl}=1.1T$, as found in lattice
calculations, when the Debye screening length became larger than the
interparticle distance or when $\alpha_s\raisebox{-.5ex}{$\stackrel{>}{\sim}$}
0.1$.  The Boltzmann equation was solved exactly to leading logarithmic order
numerically but the result only differed by less than a percent from an
analytical result obtained by a simple ansatz for the deviation from local
equilibrium, $\Phi\propto p_xp_y$.  The next to leading orders was also
calculated and found to be very important for the transport properties relevant
for quark-gluon plasmas created in relativistic heavy ion collisions and the
early universe.  For $\alpha_s\raisebox{-.5ex}{$\stackrel{>}{\sim}$} 0.1$ we
find $\eta_i= C_{i,1}T^3/\alpha_s^2\ln(C_{i,2}/\alpha_s)$ whereas for
$\alpha_s\raisebox{-.5ex}{$\stackrel{>}{\sim}$} 0.1$ we find $\eta_i=C_{i,3}
T^3/\alpha_s^2$ with coefficients $C_{i,j}$ given above.

The viscosity in degenerate plasmas of quarks, i.e., for $T\ll\mu_q$ was
calculated in \cite{deg}. Several differences were found.  In the high
temperature quark-gluon plasma the chemical potential can be ignored and the
transport processes depend on two momentum scales only, namely $T$ and $q_D\sim
gT$. In degenerate quark matter three momentum scales enter, namely $\mu_q$, T,
and $q_D\sim g\mu_q$, and the transport process depends considerably on which
of $q_D$ and $T$ is the larger. In fact for $T\ll q_D$ transverse interactions
turn out to be dominant in contrast to the high temperature quark-gluon plasma
where transverse and longitudinal interactions contribute by similar magnitude.
Furthermore, the existence of a relative sharp Fermi surface allows an almost
analytical calculation of both the leading (logarithmic) order as well as the
next order $\alpha_s^2$.

The techniques for calculating the viscosities to leading orders in the
coupling constants can be applied to other transport coefficients as well. The
leading logarithmic orders to momentum stopping, electrical conductivities and
thermal dissipation in QCD and QED plasmas have been estimated with simple
ans\"{a}tze for the distribution functions in \cite{BP3}. Based on the
experience with the viscosity studied here, we do not expect the leading
logarithmic order for these transport coefficients to decrease by much when a
full variational calculation is performed.  The next to leading logarithmic
order for these transport coefficients can also be estimated with the
experience obtained above for the viscosity.  Good estimates are obtained if
one in (\ref{qint}) replaces $q_{max}$ by the average particle momenta entering
the collision integral for the relevant transport process and $q_{min}$ by
$\sim q_D$.

A few transport coefficients are, however, different.  The second viscosity
$\zeta$ is zero for a gas of massless relativistic particles \cite{Ga} and one
cannot define a thermal conductivity in a plasma of zero baryon number. One
can, however, consider thermal dissipation processes \cite{BP3} where the
leading orders also can be calculated with the above methods. The effective
soft cutoff will, however, be different for thermal dissipation processes as
described in \cite{deg} because the transport of energy introduce dependences
on $\omega$ which also is present in the transverse screening,
$\Pi_T(\omega/q)$.

All the transport processes discussed above depend only on momentum scales from
the typical particle momentum, $q_{max}\sim T$ down to the Debye screening
wavenumber $q_{min}\sim q_D\sim gT$ which also is the momentum scale for
dynamical screening. There is, however, a shorter momentum scale of order the
magnetic mass, $m_{mag}\sim g^2T$, at which perturbative ideas of the
quark-gluon plasma fails \cite{Linde}. As shown in \cite{HP} the quark and
gluon quasiparticle decay rates depend on this infrared cutoff, $m_{mag}$.
Furthermore, recent studies \cite{color} find that the color diffusion and
conductivity also depend on this cutoff and therefore the rate of color
relaxation is a factor $1/\alpha_s$ larger than Eq. (\ref{tintro}).

\acknowledgments

This work was supported by DOE grant No. DE-AC03-76SF00098,
NSF grant No. PHY 89-21025 and the Danish
Natural Science Research Council. Discussions with Gordon Baym and Chris
Pethick are gratefully acknowledged.

\appendix{Exact Variational Calculation to leading logarithmic order}

In this appendix we solve the Boltzmann equation and find the
deviation from local equilibrium, $\Phi$, by a variational treatment
of Eq. (\ref{eta}).

For a general function $\Phi=\hat{\bf p}_x\hat{\bf p}_yf(p/T)$ we have
\begin{eqnarray}
 \Phi_1 &+&\Phi_2 -\Phi_3-\Phi_4 = \hat{\bf p}_{1,x}\hat{\bf p}_{1,y} f(p_1)
           + \hat{\bf p}_{2,x}\hat{\bf p}_{2,y} f(p_2) \nonumber\\
           && \hspace{2cm} -\hat{\bf p}_{3,x}\hat{\bf p}_{3,y}f(p_3)
           - \hat{\bf p}_{4,x}\hat{\bf p}_{4,y}f(p_4) \nonumber\\
   &=& -(\hat{\bf q}_x\hat{\bf p}_y+\hat{\bf q}_y\hat{\bf p}_x)f(p)-\mu
\hat{\bf p}_x\hat{\bf p}_y f_1(p) \nonumber\\
   & & +(\hat{\bf q}_x\hat{\bf p}_y'+\hat{\bf q}_y\hat{\bf p}_x')f(p') +
\mu \hat{\bf p}_x'\hat{\bf p}_y' f_1(p')  \, ,
\end{eqnarray}
where we have changed notation to
${\bf p}={\bf p}_1+{\bf q}/2={\bf p}_3-{\bf q}/2$ and
${\bf p}'={\bf p}_2-{\bf q}/2={\bf p}_4+{\bf q}/2$. We have used that
the energy conserving $\delta$-functions of (\ref{PIP2}) implies
$\hat{\bf p}\hat{\bf q}=\hat{\bf p}'\hat{\bf p}=\mu$.
Furthermore, have defined the function
\begin{eqnarray}
   f_1(p)=p^3 d(f/p^2)/dp=pf'-2f \, ,
\end{eqnarray}
that vanishes when $f\propto p^2$ which was the case for the
ansatz used in section V.

For small momentum transfer the matrix element (\ref{Mgg})
depends only on energy and momentum transfer and the azimuthal angle
$\phi$. The $\delta$-functions
of (\ref{PIP2}) taking care of energy conservation fixes the polar angles
$\theta_1$ and $\theta_2$ with respect to ${\bf q}$.
Thus all the angular integrals for fixed $\mu$ and $\phi$ reduces to
rotating the three vectors ${\bf q}$, ${\bf p}$ and ${\bf p}'$ over all
Euler angles keeping  them fixed relatively to each other.
Only  $(\Phi_1+\Phi_2-\Phi_3-\Phi_4)^2$ depends on the Euler angles
and the
integration or averaging over the three Eulerian angles, while
keeping the relative positions of the vectors ${\bf q}$, ${\bf p}$ and
${\bf p}'$ fixed, i.e. keeping $\mu$ and $\phi$ fixed, gives
\begin{eqnarray}
  \overline{(\hat{\bf q}_x\hat{\bf p}_y+\hat{\bf q}_y\hat{\bf p}_x)^2} =
  \overline{(\hat{\bf q}_x\hat{\bf p}_y'+\hat{\bf q}_y\hat{\bf p}_x')^2} &=&
\frac{3+\mu^2}{15}  \, ,
\end{eqnarray}
\begin{eqnarray}
  \overline{(\hat{\bf p}_x\hat{\bf p}_y)^2}  =
\overline{(\hat{\bf p}_x'\hat{\bf p}_y')^2} &=&
                 \frac{1}{15} \, ,\\
  \overline{(\hat{\bf q}_x\hat{\bf p}_y+\hat{\bf q}_y\hat{\bf p}_x)
(\hat{\bf p}_x\hat{\bf p}_y\mu)} &=&\frac{2}{15}\mu^2 \, ,\\
  \overline{(\hat{\bf q}_x\hat{\bf p}_y'+\hat{\bf q}_y\hat{\bf p}_x')
(\hat{\bf p}_x'\hat{\bf p}_y'\mu)} &=&\frac{2}{15}\mu^2 \, ,\\
  \overline{(\hat{\bf q}_x\hat{\bf p}_y+\hat{\bf q}_y\hat{\bf p}_x)
(\hat{\bf q}_x\hat{\bf p}_y'+\hat{\bf q}_y\hat{\bf p}_x')} &=&
                 \frac{1}{15}(3\hat{\bf p}\hat{\bf p}'+\mu^2)   , \label{E0}
\end{eqnarray}
\begin{eqnarray}
  \overline{(\hat{\bf q}_x\hat{\bf p}_y+\hat{\bf q}_y\hat{\bf p}_x)
(\hat{\bf q}_y'\hat{\bf p}_x')} &=&
  \overline{(\hat{\bf q}_x'\hat{\bf p}_y'+\hat{\bf q}_y'\hat{\bf p}_x')
(\hat{\bf q}_y\hat{\bf p}_x)} \nonumber\\
  &=&    \frac{\mu^2 }{15}(3\hat{\bf p}\hat{\bf p}'-1)   \, ,
\end{eqnarray}
\begin{eqnarray}
  \overline{(\hat{\bf q}_y\hat{\bf p}_x)(\hat{\bf p}_x'\hat{\bf p}_y'\mu^2)}
   &=& \frac{\mu^2 }{30}(3(\hat{\bf p}\hat{\bf p}')^2-1)  \, ,  \label{E}
\end{eqnarray}
where $\hat{\bf p}={\bf p}/\epsilon_p$
and $\hat{\bf p}'={\bf p}'/\epsilon_p'$.
Since we assume that the plasma temperature is much larger than any of the
particle masses, the particles are relativistic and
$\hat{\bf p}$, $\hat{\bf p}'$
and $\hat{\bf q}$ are unit vectors.
The vector product of $\hat{\bf p}$ and $\hat{\bf p}$ is most useful in terms
of $\mu$ and $\phi$ (see Fig. (\ref{Geometry})
\begin{eqnarray}
    \hat{\bf p}\hat{\bf p}' = \mu^2 + (1-\mu^2)\cos\phi \, .
\end{eqnarray}

Next we integrate over $\mu$ and $\phi$. The $\mu$ integration averages
$\mu^2$ to $1/3$ whereas the $\phi$ integration is weighted by a factor
$(1-\cos\phi)^2$ from the matrix elements. Thus
we find that  (\ref{E0}-\ref{E}) vanishes whereby all
combinations mixing $p_1$ and $p_2$ very conveniently disappear.
After averaging over both Euler angles and $\mu$ and $\phi$
we obtain
\begin{eqnarray}
   \langle\overline{(\Phi_1+\Phi_2-\Phi_3-\Phi_4)^2}\rangle &=&
   \frac{q^2}{p^2}\frac{1}{15}(10f^2+f_1^2+4ff_1) \nonumber\\
   &=&  \frac{q^2}{p^2}\frac{2}{5}(f^2+\frac{1}{6}p^2f'^2) \, .
\end{eqnarray}

Let us first consider the pure gluon plasma for which (\ref{I3}) gives
\begin{eqnarray}
    \langle\Phi|I|\Phi\rangle &=& \frac{8\pi}{15} g^4 \ln(T/q_D) T^3
     \nonumber\\
    &\times&  \int_0^\infty \left[ f^2+\frac{1}{6}p^2f'^2 \right]
    \left(-\frac{\partial n}{\partial\epsilon_p}\right) \, dp
    \, , \label{I6}
\end{eqnarray}
where $n=(\exp(p/T)-1)^{-1}$ is the gluon distribution function. Since
\begin{eqnarray}
    \langle\Phi|X\rangle = \frac{8}{15\pi^2} \int f p^3
    \left(-\frac{\partial n}{\partial\epsilon_p}\right) \, dp \, ,
\end{eqnarray}
we find from (\ref{eta})
\begin{eqnarray}
    \eta = \frac{15}{8\pi g^4}\frac{T^3}{\ln(T/q_D)} \,
   \frac{\displaystyle{\left( \int_0^\infty fn' x^3dx \right)^2} }
     { \displaystyle{ \int_0^\infty (f^2+\frac{1}{6}x^2f'^2)n'dx } }
     \, , \label{eta5}
\end{eqnarray}
where $x=p/T$.
As mentioned above, the function $f(x)$ is determined by minimizing
(\ref{eta5}).
A functional variation with respect to $f$
results in a second order inhomogeneous differential equation for $f$
\begin{eqnarray}
   f''+(\frac{2}{x}+\frac{n''}{n'})f' -\frac{6}{x^2}f= -\tilde{C}x
   \, ,\label{DIF}
\end{eqnarray}
where $n''/n'=-(1+2n)$. $\tilde{C}$ is an arbitrary
constant that, by rescaling $f$, can be chosen as $\tilde{C}=2$
for convenience.

For $x\gg 1$ we can approximate $n\simeq 0$ and so we find the solution to
(\ref{DIF}), that does not increase exponentially for $x\to\infty$, to be
\begin{eqnarray}
   f(x)=x^2 \, , \quad x\gg 1\, . \label{fi}
\end{eqnarray}
For $x\ll 1$ we can approximate $n\simeq 1/x$ and the solution to (\ref{DIF})
that is finite at the origin is
\begin{eqnarray}
   f(x)= x^3(C-\frac{1}{5}\ln x) \, , \quad  0\le x\ll 1 \, ,
   \label{f0}
\end{eqnarray}
where $C\simeq 0.7$ is a constant that can only be determined by finding the
full solution to (\ref{DIF}) and matching it to (\ref{f0}).
This is done by a numerical Runge-Kutta integration
and the result is shown in Fig. (\ref{ffig}). The viscosity is now found by
inserting $f$ in (\ref{eta5}). The exact value for $\eta$ thus obtained
is only 0.523\% less than the approximate
value, $\eta_{gg}$, of Eq. (\ref{etag1}).
Since the exact value is a variational minimum, it has to be smaller than
that of (\ref{etagg}) is only slightly less because $f\simeq x^2$ for
large as just as the ansatz of (\ref{fa}) and $f$ is mainly sampled over
values of $x=p/T\gg 1$ because the integrals over
$p_1$ and $p_2$
in (\ref{I3}) have powers $\sim p^4$ to $\sim p^5$ times $n_p(1+n_p)$.
In Ref. \cite{BP1} a variational calculation with trial functions
$f(p)\propto p^\nu$ lead to a minimal viscosity for $\nu=2.104$.
This result is close to the quadratic power
of (\ref{fi}) but tends slightly towards the asymptotic form of (\ref{f0})
(see also Fig. (\ref{ffig}). It has almost the same slope and curvature
as the exact solution around $p=5T$ (note that the absolute value is
unimportant since it cancels in the viscosity).
The corresponding viscosity was 0.364\% smaller than that of (\ref{etagg}),
i.e., in between the exact result and the ansatz $f\propto x^2$.

The above analysis was restricted to a pure glue plasma. As mentioned
above the distribution functions are weighted with several powers of
momentum and we do not find much difference between fermions and bosons.
Therefore the deviation from local equilibrium for quarks
will not be much different from gluons and we can be confident that
the ansatz, $\Phi\propto p_xp_y$, of Eq. (\ref{fa}) will be a good
approximation for quarks as well accurate within less than a percent.

\appendix{Soft and Hard contributions}

The essential contribution to $\langle\Phi|I|\Phi\rangle$ is the integral
\begin{eqnarray}
    Q\left(\frac{q_{max}}{q_{min}}\right)
    &=& \frac{1}{3} \int_{-1}^{1} d\mu \int_0^{q_{max}}
    q^3\, dq [ \frac{1}{|q^2+\Pi_L(\mu)|^2} \nonumber\\
    && \hspace{.5cm} + \frac{1/2}{|q^2+\Pi_T(\mu)/(1-\mu^2)|^2} ]
    \label{Q} \, .
\end{eqnarray}
For dimensional reasons the function $Q$ can only depend on the ratio of
$q_{max}$ to the momentum scale, $q_{min}$, which is provided by the
screening. For Debye and dynamical screening $q_{min}\sim q_D$ whereas
lattice calculations of strongly interacting plasmas give
$q_{min}\sim m_{pl}=1.1T$.

As described in connection with screening non-perturbative effects
become important when $q_D\raisebox{-.5ex}{$\stackrel{>}{\sim}$} T$ which
corresponds to $\alpha_s\raisebox{-.5ex}{$\stackrel{<}{\sim}$} 0.1$.
We shall treat the two limits separately starting with the weakly interaction
plasmas for which the gluon self energies, $\Pi_{L,T}(\mu)$, are given by
Eqs. (\ref{pl}) and (\ref{pt}).
It is straight forward to calculate $Q$ numerically and the result will be
given below, but let us first make a simple analytical estimate.
The main contribution to this integral can be obtained by
including the leading terms in the self energies (\ref{pl},\ref{pt})
\begin{eqnarray}
    \Pi_L(q,\omega) &\simeq& q_D^2  \, ,  \label{pla}\\
    \Pi_T(q,\omega) &\simeq& i\frac{\pi}{4}\mu q_D^2
      \label{pta} \, .
\end{eqnarray}
Thus we find for (\ref{Q})
\begin{eqnarray}
    Q \left(\frac{q_{max}}{q_{min}}\right)
  &=& \frac{1}{3}[ \ln\left(1+\frac{q^2_{max}}{q_D^2}\right)
        - \frac{q^2_{max}}{q_D^2+q^2_{max}}   \nonumber\\
  && +\frac{1}{4}\ln\left(1+\frac{q_{max}^4}{q_D^4}(\frac{4}{\pi})^2\right)
   \nonumber\\
  &&   +\frac{2}{\pi}\frac{q_{max}^2}{q_D^2}
      Arctg\left(\frac{\pi}{4}\frac{q_D^2}{q_{max}^2}\right)  ]
      \, . \label{qin}
\end{eqnarray}
Expanding in the limit $q_{max}\gg q_D$ or equivalently for
small $\alpha_s$ we obtain the leading orders up to $\alpha_s^2$
in the coupling constant
\begin{eqnarray}
  Q \left(\frac{q_{max}}{q_{min}}\right)
  &\simeq& \ln\left(\frac{q_{max}}{q_{min}}\right)
  \, ,\quad q_{max}\gg q_D \, ,   \label{qint}
\end{eqnarray}
where
\begin{eqnarray}
   q_{min} = q_D \exp\left\{\frac{1}{6}\left(1-\ln\frac{4}{\pi}\right)\right\}
          \simeq 1.13 q_D \, . \label{qmin}
\end{eqnarray}
The two terms in (\ref{qint}) corresponding to $\ln q_{max}$ and $\ln q_{min}$
are often referred to as ``hard" and ``soft" contributions in the literature
\cite{Thoma}.

A numerical evaluation of (\ref{Q}) with $\Pi_{L,T}$ given by
Eqs. (\ref{pl}) and (\ref{pt}) instead of (\ref{pla}) and (\ref{pta})
gives a slightly larger value for
the effective minimum momentum transfer
\begin{eqnarray}
   q_{min}=1.26 q_D\, ,
\end{eqnarray}
because the additional terms in $\Pi_{L,T}$ lead to some additional
screening besides the Debye screening and Landau damping of (\ref{pla})
and (\ref{pta}). This effective
cutoff is determined by the screening only and is therefore
the same for gluon-gluon, quark-gluon and quark-quark scattering.
Whereas $q_{min}$
may serve as an effective ``cutoff" of small momentum transfers interactions,
it is not a parameter put in by hand as discussed in \cite{Heinz}.
Contrarily, it is caused and determined by Debye and dynamical screening.

If the transverse interactions are assumed to be Debye screened like the
longitudinal ones, i.e., $\Pi_T=q_D^2$, then the result would have been
$q_{min}=q_D\exp(0.5)=1.65 q_D$. This is because dynamical screening of Eq.
(\ref{pta}) is less effective than the Debye screening of (\ref{pla}) and thus
results in a smaller $q_{min}$.

It is convenient to express the results in
terms of $\alpha_s$. In weakly interacting plasmas we find
\begin{eqnarray}
  Q\left(\frac{q_{max}}{q_{min}}\right) &\simeq&
\ln\left(\frac{q_{max}}{1.26q_D}\right) =
 \nonumber\\ &=& \frac{1}{2}
\ln\left(\frac{(q_{max}/T)^2}{4\pi\alpha_s(1+N_f/6)}\right)
 ,\quad \alpha_s\raisebox{-.5ex}{$\stackrel{<}{\sim}$} 0.1 . \label{QDD}
\end{eqnarray}

The upper effective cutoff $q_{max}$ is provided by the quark and
gluon distribution functions as discussed in connection with Eq.
(\ref{PIP3}) and it therefore varies somewhat with particle type.
Because Bose
distribution functions emphasize smaller momenta than Fermi ones,
$q_{max}$ is larger for quarks.
We find $q_{max}^{(gg)}=3.0T$, $q_{max}^{(gq)}=3.8T$, and
$q_{max}^{(qq)}=4.8T$ for gluon-gluon,
quark-gluon, and quark-quark scattering respectively.
The lower effective cutoff $q_{min}$ is, however, the same for the
three cases because it only
depends on the screening in the gluon propagator.
Furthermore, we find that the extra terms in the matrix elements of
(\ref{Mgg}), (\ref{Mgq}) and (\ref{Mqq}) besides the $t^{-2}$ part
do not contribute much since they have varying signs
and turn out to be partially cancelling.
Thus the constants within the logarithms of
(\ref{etag}) and (\ref{etaq}) just reflects the different $q_{max}$ for
gluon-gluon, gluon-quark and quark-quark scattering.

Lacking screening of transverse interactions in the static limit, it has
often been assumed that some mechanism like Debye screening might lead
to screening of transverse interactions as well, i.e. $m_{pl}=q_D$.
Recently, lattice gauge calculations of QCD plasmas have found effective
screening masses of order $m_{pl}\simeq 1.1T$ near the phase transition
point, $T_c\simeq 180$MeV. In both cases it is thus assumed that
\begin{eqnarray}
   \Pi_L=\Pi_T/(1-\mu^2)= m_{pl} \,
\end{eqnarray}
in (\ref{Q}) which leads to
\begin{eqnarray}
    Q \left(\frac{q_{max}}{q_{min}}\right)
  &=& \frac{1}{2}\left[ \ln\left(1+\frac{q^2_{max}}{m_{pl}^2}\right)
        - \frac{q^2_{max}}{m_{pl}^2+q^2_{max}} \right]  \, .\nonumber\\
  && \label{Qmpl}
\end{eqnarray}
With $q_{max}^{(gg)}=3.0T$, $q_{max}^{(gq)}=3.8T$, $q_{max}^{(qq)}=4.8T$ and
$m_{pl}=1.1T$ we find $Q(q_{max}^{(gg)}/q_{min})=0.626$,
$Q(q_{max}^{(gq)}/q_{min})=0.819$ and $Q(q_{max}^{(qq)}/q_{min})=1.024$.
These values enter the $\alpha_s\raisebox{-.5ex}{$\stackrel{>}{\sim}$} 0.1$
expressions of Eqs.
(\ref{etag1}) and (\ref{etaq1}).

\newpage
\figure{Feynman diagram for gluon-gluon scattering in the t-channel.
The lines in the loops can be either quark or gluon propagators.
\label{bubble}}

\figure{The collision geometry. For small momentum transfer, $q\ll p_1,p_2$,
energy and momentum conservation requires
$\cos\theta_1=\cos\theta_2=\omega/q$. \label{Geometry}}

\figure{The gluon viscosity for $N_f=3$ assuming Debye and dynamical
screening (dashed curve), a constant screening mass $m_{pl}=1.1T$
(dashed-dotted curve) and the minimal one (full curve). \label{figetag}}

\figure{The quark, gluon and total viscosities for $N_f=3$. \label{figeta2}}

\figure{The quark, gluon and total viscosities for $N_f=2$. \label{figeta3}}

\figure{The function $f/x^2$ as determined by (\ref{DIF}). Also shown are
the limits of (\ref{fi}), (\ref{f0}) and the simple ansatz
$f=x^2$. \label{ffig}}

\end{document}